\documentclass[useAMS,twocolumn,usenatbib]{mn2e}
\usepackage{graphicx}

\usepackage{amssymb}
\usepackage{amsmath}
\usepackage{bm}

\def\OutputDriver{pdftex}

\usepackage[\OutputDriver,hyperindex]{hyperref}
\hypersetup{
breaklinks = {true},
colorlinks = {false},
linkcolor={black},
pdfpagemode = {None}, 
pdfborder = {0 0 1},
pdftitle = {Strong lensing optical depths in a $\Lambda$CDM universe II: the influence of the stellar mass in galaxies},
pdfsubject = {},
pdfauthor = {S. Hilbert, S.D.M. White, J. Hartlap \& P. Schneider},
pdfkeywords = {gravitational lensing, dark matter, large-scale structure of the Universe, galaxies: general, cosmology: theory, methods: numerical}
}

\newcommand{\mrm}[1]{\mathrm{#1}}
\newcommand{\mbb}[1]{\mathbb{#1}}

\newcommand{\vect}[1]{\boldsymbol{#1}}
\newcommand{\matr}[1]{\boldsymbol{\mathsf{#1}}}

\newcommand{\totder}[2]{\frac{\mrm{d} #1}{\mrm{d} #2}}
\newcommand{\p}{\partial}

\newcommand{\diff}[2][]{\mrm{d}^{#1}{#2}\,}

\newcommand{\tr}{\mrm{tr}\,}

\newcommand{\pdf}{\mrm{pdf}}

\newcommand{\arcsect}{\ensuremath{\mathrm{arcsec}}}
\newcommand{\kpc}{\ensuremath{\mathrm{kpc}}}
\newcommand{\Mpc}{\ensuremath{\mathrm{Mpc}}}

\newcommand{\Msolar}{\ensuremath{\mathrm{M}_\odot}}

\newcommand{\IIorIII}{\ensuremath{{\mrm{II}\vee\mrm{III}}}}

\newcommand{\zS}{z^\mrm{S}}
\newcommand{\zL}{z^\mrm{L}}

\newcommand{\Mstellar}{\ensuremath{M_\mrm{stellar}}}
\newcommand{\Mdisc}{\ensuremath{M_\mrm{disc}}}
\newcommand{\Mbulge}{\ensuremath{M_\mrm{bulge}}}

\newcommand{\Sdisc}{\ensuremath{\Sigma_\mrm{disc}}}
\newcommand{\Sbulge}{\ensuremath{\Sigma_\mrm{bulge}}}
\newcommand{\rsdisc}{\ensuremath{r_\mrm{s\,disc}}}

\newcommand{\reffbulge}{\ensuremath{r_\mrm{e\,bulge}}}
\newcommand{\rEinstein}{r_\mrm{E}}

\newcommand\aj{{AJ}} 
\newcommand\apj{{ApJ}} 
\newcommand\apjl{{ApJ}} 
\newcommand\apjs{{ApJS}} 
\newcommand\aap{{A\&A}} 
\newcommand\aaps{{A\&AS}} 
\newcommand\mnras{{MNRAS}} 
\newcommand\nat{{Nature}} 

\defcitealias{HilbertEtal2007_StrongLensing}{paper~I}

\begin{document}

\title[Strong lensing optical depths... II]{Strong lensing optical depths in a $\Lambda$CDM universe II: the influence of the stellar mass in galaxies}
\author[S. Hilbert, S.D.M. White, J. Hartlap \& P. Schneider]{
Stefan Hilbert,$^1$\thanks{\texttt{hilbert@mpa-garching.mpg.de}}
Simon D.~M. White,$^1$
Jan Hartlap,$^2$
and Peter Schneider$^2$
\\$^1$Max-Planck-Institut f{\"u}r Astrophysik,Karl-Schwarzschild-Stra{\ss}e 1, D-85741, Garching, Germany
\\$^2$Argelander-Institut f{\"u}r Astronomie, Auf dem H{\"u}gel 71, D-53121 Bonn, Germany
}

\date{\today}

\maketitle

\begin{abstract}
We investigate how strong gravitational lensing in the concordance $\Lambda$CDM cosmology is affected by the stellar mass in galaxies. We extend our previous studies, based on ray-tracing through the Millennium Simulation, by including the stellar components predicted by galaxy formation models. We find that the inclusion of these components greatly enhances the probability for strong lensing compared to a `dark matter only' universe. The identification of the `lenses' associated with strong-lensing events reveals that the stellar mass of galaxies (i) significantly enhances the strong-lensing cross-sections of group and cluster halos, and (ii) gives rise to strong lensing in smaller halos, which would not produce noticeable effects in the absence of the stars. Even if we consider only image splittings $\gtrsim10\,\arcsect$, the luminous matter can enhance the strong-lensing optical depths by up to a factor of 2.
\end{abstract}


\section{Introduction}
\label{sec:Introduction}

The $\Lambda$CDM model, the current standard model of cosmological structure formation, is based on a flat universe with cold dark matter and a cosmological constant. It has been shown to fit a wide range of observations, including the properties of galaxies, their clustering, the accelerated expansion inferred from the apparent luminosity of distant type Ia supernovae, the structure of the high-redshift intergalactic medium, and temperature fluctuations in the cosmic microwave background~\citep{SpergelEtal2007_WMAP_3rdYear_Data}.

Further tests and constraints on the parameters of the $\Lambda$CDM model are obtained from measurements of gravitational lensing effects, which were first discovered through multiple images of distant quasars~\citep{WalshCarswellWeymann1979} and highly distorted images of distant galaxies at optical~\citep{LyndsPetrosian1986,SoucailEtal1987} and radio~\citep{HewittEtal1988} wavelengths. The results of the first optical surveys for multiply imaged quasars \citep{CramptonETal1992_HRCam_Survey,YeeFilippenkoTang1993,MaozEtal1993_HST_Snapshot_Survey,SurdejEtal1993} were used to constrain the value of the cosmological constant \citep{MaozRix1993,Kochanek1993_II,Kochanek1996}. Later quasar strong-lensing surveys \citep{InadaEtal2008} have been used to constrain its possible evolution \citep{OguriEtal2008}.
Many recent observations of gravitational-lensing effects around galaxies \citep[e.g.][]{MandelbaumEtal2006_Galaxies,SimonEtal2007}, and in and around galaxy clusters \citep[e.g.][]{MandelbaumEtal2006_GroupsAndClusters,CloweEtal2006,NatarajanDeLuciaSpringel2007,ComerfordEtal2006,MasseyEtal2007_cosmic_scaffolding} are well explained by the dark-matter structures predicted from the $\Lambda$CDM model. Surveys measuring the weak lensing effects of the matter distribution as a whole \citep[e.g.][]{SemboloniEtal2006,HoekstraEtal2006,SimonEtal2007,MasseyEtal2007_3D_WL,BenjaminEtal2007,FuEtal2008} are particularly promising for further constraining the parameters of the $\Lambda$CDM model. An open question is whether the observed frequency of giant arcs \citep{LuppinoEtal1999,ZaritskyGonzalez2003,GladdersEtal2003} is consistent with predictions based on the $\Lambda$CDM model with parameters favoured by other observations ~\citep[e.g.][]{BartelmannEtal1998,OguriLeeSuto2003,DalalHolderHennawi2004,WambsganssBodeOstriker2004,LiEtal2006,MeneghettiEtal2007}.

The efforts currently underway to improve the measurement of lensing effects need to be matched by a comparable improvement in the theoretical predictions. According to the $\Lambda$CDM model, most of the matter in our Universe is dark. The baryonic matter contributes significantly, however, to the inner regions of galaxies and clusters, which is where strong lensing is observed. Most theoretical studies of strong lensing by galaxies  \citep[e.g.]{TurnerOstrikerGott1984,MaozRix1993,MoellerBlain2001,HutererKeetonMa2005,Oguri2006,Chae2007,MoellerKitzbichlerNatarajan2007,OguriEtal2008} use analytic profiles to model both the luminous and the dark components of the lenses \citep[see][for a review]{Kochanek2006_SaasFee33_Part2}. Many studies of strong lensing by galaxy clusters use profiles obtained from $N$-body simulations for the dark matter, but neglect the luminous matter \citep[e.g.][]{BartelmannEtal1998,WambsganssBodeOstriker2004,LiEtal2006}. There are, however, also studies of the effect of galaxies on giant-arc probabilities in clusters. \cite{MeneghettiEtal2000}, for example, placed galaxies randomly into simulated dark-matter clusters and concluded that, although the galaxies change certain aspects of cluster lensing, they do not have a strong effect on the formation of giant arcs. \cite{FloresMallerPrimack2000} used analytic profiles for the cluster halo and the cluster galaxies and found that the galaxies do only mildly increase the cluster's cross-section for giant arcs. \cite{MeneghettiBartelmannMoscardini2003} and \cite{DalalHolderHennawi2004} studied the effect of a large central galaxy in a cluster, and found that even a very massive central galaxy does not greatly affect the giant-arc cross-section. \citet{WambsganssOstrikerBode2008} observed moderate effects of baryons on the frequencies of giant arcs and large image splittings by using a simple description of baryon condensation to place galaxies in simulated dark-matter halos.

\citet{PuchweinEtal2005} and \citet{RozoEtal2006_astroph} incorporated a treatment of the baryonic component into cluster formation simulations and studied its influence on giant-arc probabilities. Such simulations currently have difficulty producing a galaxy population which matches observation, so their results are not simple to interpret. Until this problem is overcome, a hybrid approach that embeds a semi-analytic treatment of galaxy formation within an $N$-body simulation of dark-matter evolution~\citep{SpringelEtal2001_SUBFIND, SpringelEtal2005_Millennium} appears the most realistic way to address these issues.

In earlier work \citep[][ \citetalias{HilbertEtal2007_StrongLensing} in the following]{HilbertEtal2007_StrongLensing}, we studied the statistics of strong lensing by shooting random rays through a series of lens planes created from the Millennium Simulation~\citep{SpringelEtal2005_Millennium}. This very large $N$-body simulation of cosmological structure formation did not explicitly include gas physics such as radiative cooling and star formation, and the results presented in \citetalias{HilbertEtal2007_StrongLensing} did not account for the effects of the stellar components of galaxies.

In this paper, we extend the work of \citetalias{HilbertEtal2007_StrongLensing} to include the gravitational effects of the stars in galaxies, as inferred from semi-analytic galaxy-formation models implemented within the evolving dark-matter distribution of the Millennium Simulation~\citep{SpringelEtal2005_Millennium,CrotonEtal2006,DeLuciaEtal2006,DeLuciaBlaizot2007}. These models, which have been adjusted to be consistent with a large number of observations, couple star formation in the galaxies directly to the properties of the underlying dark matter. They currently provide the most accurate way to simulate how the dark matter distribution is populated with galaxies.

The Millennium Simulation has higher resolution and a much larger volume than simulations used in previous studies of the effects of galaxies on strong lensing. On scales above the resolution limit, the simulation provides a more realistic matter distribution than analytic models. Furthermore, the galaxy model we use provides the stellar mass of the galaxies, computed from the assembly history of their dark-matter halos, as well as their positions with respect to the dark matter in the simulation. This is a considerable improvement over the simpler recipes used to place galaxies into dark matter halos in previous work \citep[e.g. by][]{MeneghettiEtal2000,MeneghettiBartelmannMoscardini2003,DalalHolderHennawi2004,WambsganssOstrikerBode2008}. Thus we hope to obtain more accurate results, particularly in those regions where both the luminous and the dark matter are important for lensing. This includes the inner few arcseconds of most lenses, regions which dominate many surveys for strong lensing \citep{MyersEtal2003,OguriEtal2006,BoltonEtal2006,WillisEtal2006,CabanacEtal2007}.

Our paper is organised as follows. In Sec.~\ref{sec:methods}, we summarise the main aspects of our method for shooting a representative ray sample through the Millennium Simulation, and we describe how we incorporate the lensing effects of the stellar mass in galaxies. In Sec.~\ref{sec:results}, we present results for the magnification distribution, for strong-lensing optical depths, and for lensing cross-sections as a function of lens halo mass and projected distance between lens centre and image. In particular, we compare the results obtained for dark matter alone to those obtained when the stellar mass of galaxies is also included. The paper concludes with a summary and outlook in Sec.~\ref{sec:summary}.

\section{Simulation methods}
\label{sec:methods}
Our approach for simulating gravitational lensing closely follows that of \citetalias{HilbertEtal2007_StrongLensing}. The reader is referred to that paper for a detailed description. Here, we summarise the main aspects of the method and discuss the extensions needed for the present work.

In order to calculate image distortions resulting from the gravitational deflection of light by matter inhomogeneities between the source and the observer, we use a Multiple-Lens-Plane algorithm \citep[e.g.][]{BlandfordNarayan1986,SchneiderEhlersFalco_book,SeitzSchneiderEhlers1994,JainSeljakWhite2000,PaceEtal2007}. Lens planes are introduced transverse to the line-of-sight, and matter inhomogeneities in the observer's backward light-cone are projected onto them. Light rays are traced back from the observer to their source under the assumption that the rays propagate unperturbed between lens planes, but are deflected when passing through a plane. The ray distortions (more precisely, the distortions of infinitesimally thin ray bundles) induced by the lens planes are calculated from the projected matter distribution on the planes.

\subsection{The dark-matter contribution}
\label{sec:dm_contribution}
We use the particle data of the of the Millennium Simulation~\citep{SpringelEtal2005_Millennium} to generate the dark-matter distribution on the lens planes. The Millennium Simulation assumes a flat $\Lambda$CDM universe with a matter density of $\Omega_\mrm{M}=0.25$ in terms of the critical density, a cosmological constant with $\Omega_\Lambda=0.75$, a Hubble constant $100h\,\mrm{km}\mrm{s}^{-1}\Mpc^{-1}$ with $h=0.73$, a primordial spectral index $n=1$ and a normalisation parameter $\sigma_8=0.9$ for the linear density power spectrum. The simulation followed $10^{10}$ particles of mass $m_\mrm{p}=8.6 \times 10^8h^{-1} \,\Msolar$ in a cubic region of side length $L=500h^{-1}\,\Mpc$ comoving from redshift $z=127$ to $z=0$. The effective resolution reached near the centres of dark-matter halos is comparable to the comoving force-softening length of $5h^{-1}\,\kpc$. During the run, 64 snapshots of the simulation were taken and stored on disk.

Along the line-of-sight, we place one plane for each snapshot at the corresponding distance from the observer, resulting in 44 planes for sources at redshift $\zS=5.7$ (the highest redshift we consider). For each lens plane, the particles within an oblique slice of appropriate thickness through the corresponding snapshot are projected onto a hierarchy of meshes (with a spacing of $2.5h^{-1}\,\kpc$ comoving for the finest mesh). The projected matter distribution is smoothed by an adaptive scheme to reduce shot noise from individual particles while retaining a resolution of about $5h^{-1}\,\kpc$ comoving in dense regions. Fast-Fourier-Transform (FFT) methods \citep{FrigoJohnson2005_FFTW3} are employed to calculate the dark-matter lensing potential from the projected particle distribution on the planes. The second derivatives of the potential, which quantify the dark-matter contribution to the distortion of light rays passing through the lens planes, are calculated by finite difference and bilinear interpolation.

\subsection{The stellar contribution}
\label{sec:stellar_contribution}
The Millennium Simulation did not explicitly simulate the physics of star formation. Rather this was done in post-processing by applying several semi-analytic models to halo merger trees generated from the stored output in order to follow the formation and evolution of the galaxies. In this paper we use the catalogue made publicly available by \citet{LemsonVirgo2006_astroph_SA_Galaxies}\footnote{http://www.mpa-garching.mpg.de/Millennium} based on the model by \citet{DeLuciaBlaizot2007} to obtain the main properties of the galaxies. Besides many other quantities, the catalogue provides the positions, stellar disc and bulge\footnote{Here, `bulge' means the spheroidal stellar component of a galaxy, e.g., the stellar bulge of a disc galaxy or the all stars in an elliptical galaxy with no disc.} masses, and disc radii. Here, we restrict our analysis to galaxies with $\Mstellar \geq 10^{9}h^{-1}\Msolar$, which we project onto the same set of lens planes as the simulation particles. As a test, we also tried a higher mass limit $\Mstellar \geq 10^{10}h^{-1}\Msolar$, finding rather minor changes. Clearly our adopted limit is quite sufficient to include all significant lenses, at least according to this model.

For each galaxy, we approximate the projected matter distribution of the disc component by an exponential surface density profile with comoving scale radius $\rsdisc$ and total mass $\Mdisc$ taken from the galaxy catalogue data:\footnote{
The catalogue does not contain $\Mdisc$ and $\rsdisc$ explicitly, but provides the total stellar mass $\Mstellar$, the stellar bulge mass $\Mbulge$, the physical disc size radius $R_\mrm{disc}$, and the redshift $z$. According to the underlying galaxy-formation model, $\Mdisc=\Mstellar-\Mbulge$ and $\rsdisc=R_\mrm{disc} (1+z)/3$.
}
\begin{equation}
\label{eq:disc_component}
\Sdisc(r)=\frac{\Mdisc}{2\pi\rsdisc^2}\exp\left(-\frac{r}{\rsdisc}\right).
\end{equation}
Here, $\Sdisc(r)$ denotes the comoving surface mass density at projected comoving distance $r$ on the lens plane.

The galaxy model of \citet{DeLuciaBlaizot2007} provides bulge masses $\Mbulge$, but not the bulge radii. We thus use an empirical relation to calculate the bulge radii from the bulge masses. \cite{ShenEtal2003}\footnote{See also \citet{ShenEtal2003_Erratum} for an Erratum.} studied the size distribution of $140\,000$ galaxies in the Sloan Digital Sky Survey (SDSS) and found a relation $\bar{R}_\mrm{e}\propto \Mstellar^{0.56}$ between the median physical effective radius $\bar{R}_\mrm{e}$ and stellar mass $\Mstellar$ of early-type galaxies at redshift $z\lesssim0.3$. \citet{TrujilloEtal2006} combined results of SDSS, GEMS, and FIRES to study the evolution of galaxy sizes between redshift $z=0$ and $z\approx3$, and found $R_\mrm{e}\propto(1+z)^{-0.45}$ for the mean radius $R_\mrm{e}$ of early types at fixed stellar mass. For the effective bulge radius $\reffbulge$ (measured in comoving units), we combine these relations into:
\begin{equation}
\label{eq:r_bulge}
\reffbulge=(1+z)^{0.55}\left(\frac{\Mbulge}{10^{10}h^{-1}\,\Msolar}\right)^{0.56}\times 0.54h^{-1}\,\kpc.
\end{equation}
The bulge component of each galaxy is then approximated by a spherical de-Vaucouleurs profile~\citep{DeVaucouleurs1948}:
\begin{equation}
\label{eq:bulge_component}
\Sbulge(r)=\frac{94.5\Mbulge}{\reffbulge^2}\exp\left[-7.67\left(\frac{r}{\reffbulge}\right)^{1/4}\right].
\end{equation}
Here, $\Sbulge(r)$ denotes the projected bulge surface mass density at projected comoving radius $r$.

For each galaxy, we use analytic expressions to calculate the ray distortions induced by the mass distributions~\eqref{eq:disc_component} and \eqref{eq:bulge_component} \citep[see, e.g.,][]{Cardone2004}. The disc and bulge contributions of all galaxies on the lens plane are then summed to obtain the stellar contribution to the distortion of light rays passing through the plane.

The profile~\eqref{eq:disc_component} for the disc component closely follows the projected mass distribution of a stellar disc seen face-on. This, of course, neglects the effects of disc inclination with respect to the line-of-sight, which on average increases the cross-section for strong lensing by isolated disc galaxies \citep[e.g.][]{MallerFloresPrimack1997,WangTurner1997,BartelmannLoeb1998,KeetonKochanek1998,MoellerBlain1998}. Similarly, the model~\eqref{eq:bulge_component} for the bulge neglects ellipticity. Furthermore, recent observations~\citep{TrujilloEtal2007} indicate a stronger evolution of galaxy size with redshift for very massive galaxies, leading to smaller radii for spheroid-like galaxies with $\Mstellar>10^{11}\,\Msolar$ at redshifts $z\gtrsim1$ than our estimate~\eqref{eq:r_bulge}. Thus our model for the stellar component of galaxies may be inaccurate for strong lensing occurring close to the centres of galaxies. For strong lensing at distances of several effective radii, however, the details of the stellar matter distribution do not influence the lensing properties very strongly.

To test the dependence of our results on the particular choice of the mass profile for the stars in galaxies, we replaced the de-Vaucouleurs profile~\eqref{eq:bulge_component} by a Hernquist profile~\citep{Hernquist1990}, and by a Plummer profile with $\Sbulge(r)\propto(\reffbulge^2+r^2)^{-2}$ and the same effective radius $\reffbulge$. For `cuspy' de-Vaucouleurs and Hernquist profiles, the optical depths and cross-sections for strong lensing by both dark and luminous matter agree within $\sim20\%$, whereas the results for the `cored' Plummer profile were noticeably smaller. Hence, our results are not very sensitive to the details of stellar mass distribution as long as we use cuspy profiles for the spheroid stellar component of the galaxies.

The particles in the Millennium Simulation are treated as collisionless particles by the simulation and are used in our calculations to obtain the dark-matter distribution on the lens planes, but they represent the \emph{total} mass in the simulated part of the universe. One could argue that the stellar mass we add to the lens planes should be removed somewhere else to balance the mass budget. Here, we refrain from doing this for two main reasons: (i) The mass in stars ($1\%$ of the total mass at $z=0$) is small compared to the mass in collapsed objects ($50\%$ at $z=0$). (ii) Although the baryons that produce the stars originate from within the halos around the galaxies, gas physics increases the dark-matter density in the inner part of the halos compared to collisionless simulations \citep[e.g.][]{BarnesWhite1984,BlumenthalEtal1986,GnedinEtal2004}. Only in the halo outskirts, which are not relevant for strong lensing, is there a net decrease of the matter density. The effects of the baryons on the dark-matter profile and their relevance for lensing~\citep[see, e.g.,][]{PuchweinEtal2005,JingEtal2006,RozoEtal2006_astroph,WambsganssOstrikerBode2008} are not the primary focus of this work, so we neglect these effects in the following.

\subsection{Sampling image distortions}
\label{sec:sampling}
The light rays traced back from the observer through the lens planes to their source define the lens mapping \mbox{$\vect{\theta}\mapsto \vect{\beta}$}, which relates the `observed' angular position $\vect{\theta}$ of a ray in the image plane $\mbb{P}^\mrm{I}$ to the 'true' angular position $\vect{\beta}$ of its source in source plane $\mbb{P}^\mrm{S}$ at a given redshift $\zS$. The distortion matrix $\matr{A}=(\p\vect{\beta}/\p\vect{\theta})$, i.e. the Jacobian of the lens mapping, can be calculated from the distortions that the ray bundles experience when passing through the lens planes.

For sufficiently small sources, the distortion matrix quantifies the image distortions induced by the deflections. In this case, the image magnification $\mu$ is given by $\mu=\left(\det{\matr{A}}\right)^{-1}$, and the eigenvalues $\lambda_1$ and $\lambda_2$ of $\matr{A}$ (w.l.o.g. $|\lambda_1|>|\lambda_2|$) determine the major-to-minor axis ratio $r=|\lambda_1|/|\lambda_2|$ of the elliptical images of circular sources \citep{SchneiderEhlersFalco_book}.

To quantify the frequency of images with a given property $p$, e.g. a large magnification or large length-to-width ratio, we define the optical depths
\begin{equation}
\label{eq:def_tau_I}
\tau^\mrm{I}_p=\frac{\int_{\mbb{P}^\mrm{I}}\diff[2]{\vect{\theta}}1_p(\vect{\theta})}{\int_{\mbb{P}^\mrm{I}}\diff[2]{\vect{\theta}}}
\end{equation}
and
\begin{equation}
\label{eq:def_tau_S}
\tau^\mrm{S}_p=
\frac{\int_{\mbb{P}^\mrm{I}}\diff[2]{\vect{\theta}}\left|\mu(\vect{\theta})\right|^{-1}1_p(\vect{\theta})}
{\int_{\mbb{P}^\mrm{I}}\diff[2]{\vect{\theta}}\left|\mu(\vect{\theta})\right|^{-1}},
\end{equation}
where $1_p(\vect{\theta})=1$ if the image at position $\vect{\theta}$ has property $p$, and $1_p(\vect{\theta})=0$ otherwise. The optical depth $\tau^\mrm{I}_p$ estimates the fraction of images with property $p$ assuming a uniform distribution of images in the image plane, as is the case, e.g., for magnitude-limited surveys and a source population with integral luminosity function inversely proportional to the threshold luminosity. The optical depth $\tau^\mrm{S}_p$ estimates the fraction of images with property $p$ assuming a uniform distribution of sources in the source plane, which is appropriate for volume-limited surveys.

In order to estimate these optical depths, the dark-matter and the stellar contributions to the distortion are calculated at 160 million randomly chosen positions on each lens plane. For every considered source redshift, the distortions from all lens planes with smaller redshift are combined at random to generate the distortion matrices for rays along $1.6\times10^8$ random lines-of-sight. Combining the distortions from different lens planes at random saves us from computing the actual path of light rays, while it allows us to sample the image distortions on a very large area under the assumption that the mass distribution projected on different lens planes is uncorrelated.

The measured fraction of rays with a certain property, e.g. a large magnification, is then used as a Monte-Carlo estimate for the corresponding optical depth $\tau^\mrm{I}$ to the chosen source redshift. The optical depth $\tau^\mrm{S}$ is obtained by additionally weighting all rays by their inverse magnification. For comparison purposes, we repeat the procedure using the dark-matter contribution alone, and using the stellar contribution alone.

\section{Results}
\label{sec:results}

\subsection{The magnification distribution}
\label{sec:magnification_distribution}

\begin{figure}
\centerline{\includegraphics[width=1\linewidth]{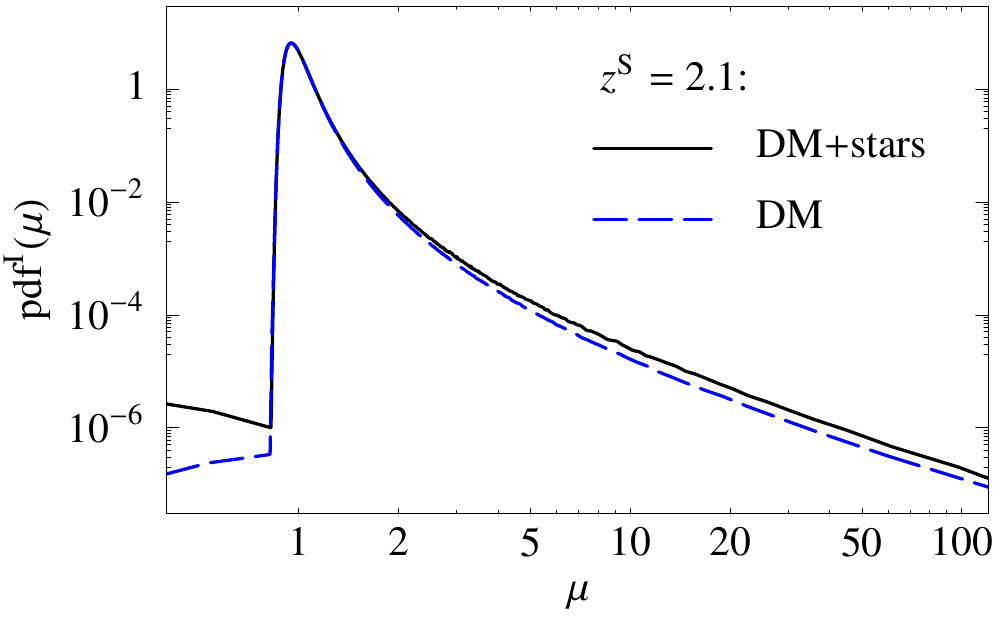}}
\caption{
\label{fig:pdf_of_mu}
The probability density \mbox{$\pdf^\mrm{I}(\mu)$} of the magnification $\mu$ for sources at redshift $\zS=2.1$. The case of lensing by dark and luminous matter (solid line) is compared to the case of dark matter alone (dashed line).
}
\end{figure}

By binning the magnifications of our random rays, we estimated the probability density function
\begin{equation}
\label{eq:def_pdf_mu}
\pdf^\mrm{I}(\mu')=\totder{}{\mu'}\tau^\mrm{I}_{\mu(\vect{\theta})\leq \mu'}.
\end{equation}
In Fig.~\ref{fig:pdf_of_mu}, the distribution including the stellar contribution is compared with the dark-matter only distribution for sources at redshift \mbox{$\zS=2.1$}. There is little difference between the two distributions at magnifications $\mu\approx1$. Only in the high-$\mu$ tail (containing the strongly focused rays) and for very low $\mu$ (containing the overfocused rays) do the two distributions differ significantly. The increased pdf at high magnifications hints at a higher optical depth for large magnifications when the effects of the luminous matter are included. Qualitatively the same behaviour is found for all considered source redshifts.

\subsection{Strong-lensing optical depths}
\label{sec:optical_depths}

\begin{figure}
\centerline{\includegraphics[width=1\linewidth]{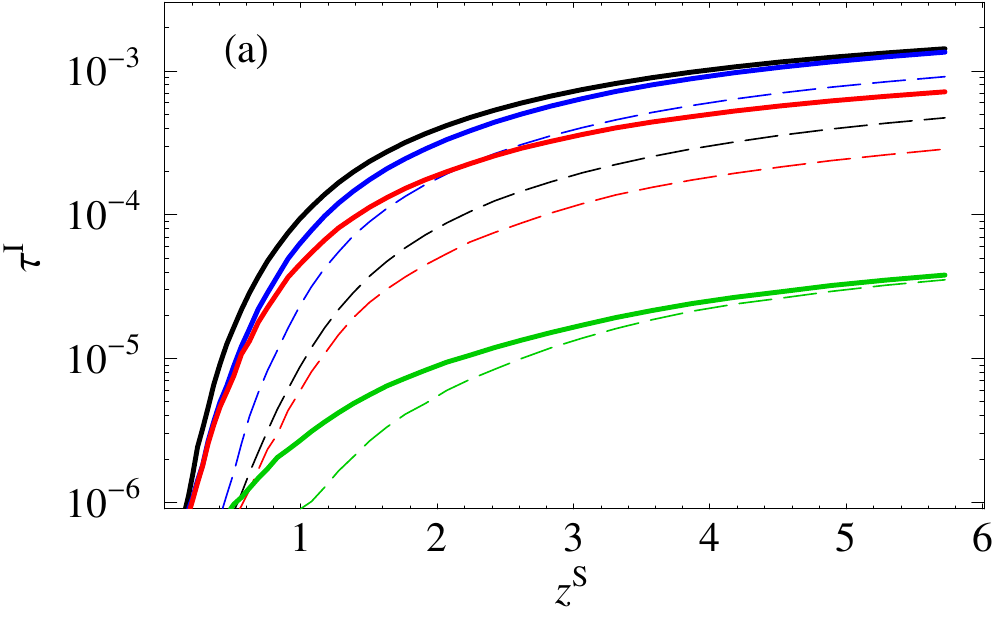}}
\centerline{\includegraphics[width=1\linewidth]{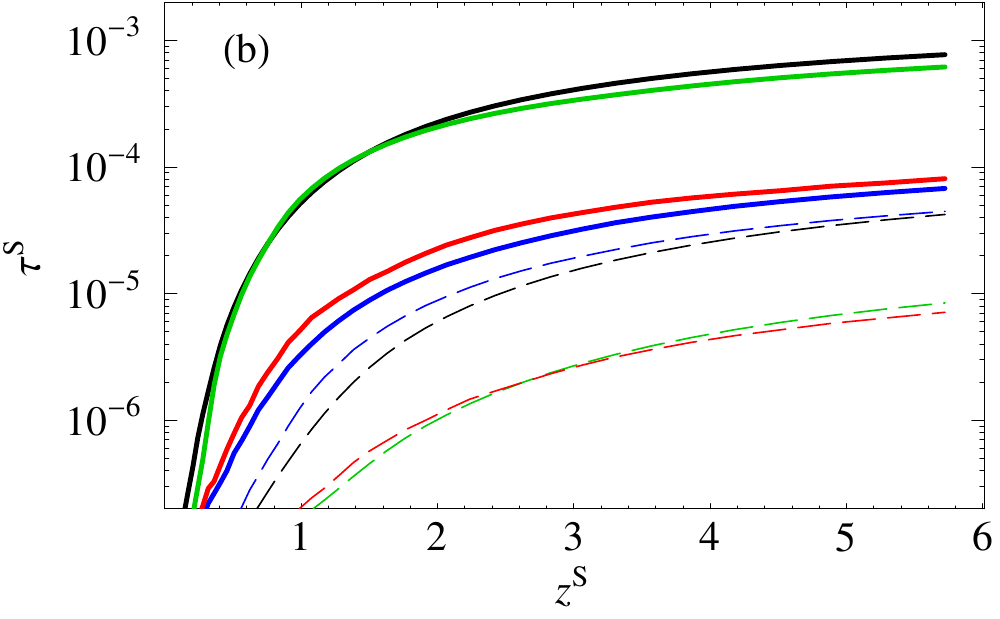}}
\caption{
\label{fig:tau}
Comparison between the optical depths caused by dark and luminous matter (solid lines) and the optical depths caused by dark matter alone (dashed lines), either assuming a uniform distribution of images in the image plane (a), or a uniform distribution of sources in the source plane (b). Shown are optical depths for images of small circular sources of type II (black lines), of type III (green), with  magnification $|\mu|>10$ (blue) and with length-to-width ratio $r>10$ (red).
}
\end{figure}

From our ray sample, we calculate
\begin{itemize}
\item the fraction of rays with \mbox{$\det\matr{A}<0$} (called \mbox{type II}),
\item the fraction with \mbox{$\det\matr{A}>0$} and \mbox{$\tr\matr{A}<0$} \mbox{(type III)},
\item the fraction with \mbox{$\det\matr{A}<0$} or \mbox{$\tr\matr{A}<0$}, i.e. the sum of the two previous classes (type \IIorIII)\footnote{In all situations relevant for this work, images of type II and III, and hence all images of type \IIorIII{}, belong to sources with multiple images.},
\item the fraction with a length-to-width ratio \mbox{$r>10$} for images of sufficiently small circular sources, and
\item the fraction with magnification \mbox{$|\mu|>10$}.
\end{itemize}
The corresponding optical depths $\tau_p^\mrm{I}(\zS)$ and $\tau_p^\mrm{S}(\zS)$ are plotted in Fig.~\ref{fig:tau} as functions of the source redshift $\zS$. The optical depths that account for the stellar mass in galaxies are significantly larger than those which neglect it. The effect is particularly large at low source redshifts, where the dark matter alone is much less efficient in producing strong lensing.

In contrast to lensing by dark matter alone, the optical depths for $|\mu|>10$ and $r>10$ are quite similar when the luminous matter is included, the optical depth $\tau^\mrm{S}_\mrm{II}$ for images of type II is no longer much smaller than $\tau^\mrm{I}_\mrm{II}$, and the optical depth $\tau^\mrm{S}_\mrm{III}$ is even larger than $\tau^\mrm{I}_\mrm{III}$. This implies that most images of type III are strongly demagnified. These are the images of multiply imaged sources that often remain undetected in observations because they are demagnified and close to the bright centre of the lens galaxy.

The optical depths for joint lensing by luminous and dark matter show similar behaviour to cored isothermal spheres.  For singular isothermal spheres, one of the two eigenvalues of the distortion matrix $\matr{A}$ is unity \citep{SchneiderEhlersFalco_book}. Hence, $|\mu|=r$, and the cross-sections for $r>10$ and $|\mu|>10$ are equal. Singular isothermal spheres do not produce images of type III. If the central singularity is replaced by a small core, the lens acquires a finite cross-section for strongly demagnified images of type III, and the cross-sections for $r>10$ and $|\mu|>10$ separate slightly. Thus, the addition of baryons apparently makes the mass profiles of strong lenses resemble isothermal spheres with small cores.

The optical depths $\tau^\mrm{S}_{\mu>10}$ for rays with magnification $\mu>10$ (shown in Fig.~\ref{fig:tau}b) are very similar to the optical depths published by \citet{WambsganssOstrikerBode2008}. The optical depths for lensing by dark matter alone agree within ten percent. However, our results show a larger enhancement due to baryons (30-60 percent) compared to their work (20-30 percent). The larger values we find may be due to the higher effective resolution of our method.

\subsection{Lens properties}
\label{sec:lens_properties}
As discussed in \citetalias{HilbertEtal2007_StrongLensing} for lensing by dark matter alone, the properties of most strongly lensed rays, i.e. rays with $\det\matr{A}<0$, $\tr\matr{A}<0$, $|\mu|>10$, or $r>10$, are predominantly caused by a single matter clump along the line of sight. This is equally the case if galactic baryons are included as lensing material. In order to find these matter clumps, which we refer to as the lenses of the rays, we use the method described in \citetalias{HilbertEtal2007_StrongLensing}: We determine for each strongly lensed ray the lens plane that is sufficient to produce the relevant property in the single-plane approximation. Depending on source redshift and the property considered, this simple criterion identifies exactly one lens plane for at least 60\% (usually $>80\%$) of the rays. The redshift of this plane is taken as the lens redshift $\zL$ for the ray.

\begin{figure}
\centerline{\includegraphics[width=1\linewidth]{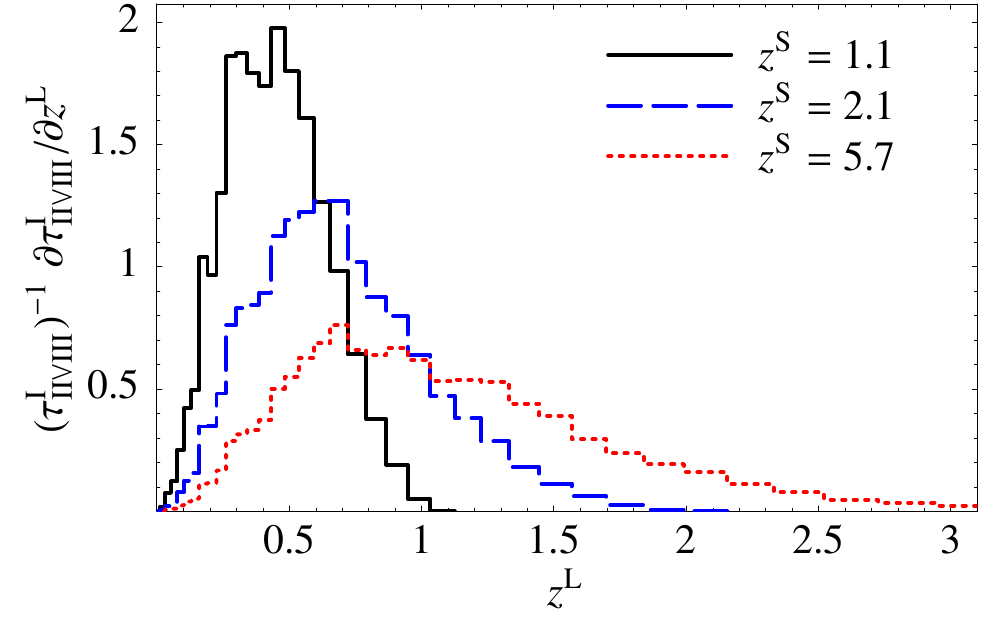}}
\caption{
\label{fig:sigma_of_zL}
The cross-section \mbox{$\partial\tau^\mrm{I}_\IIorIII/\partial z^\mrm{L}$} for rays of type \IIorIII\ as a function of lens redshift $z^\mrm{L}$ for sources at redshift \mbox{$z^\mrm{S}=1.1$} (solid line), $z^\mrm{S}=2.1$ (dashed line), and \mbox{$z^\mrm{S}=5.7$} (dotted line) considering lensing by both dark and luminous matter.
}
\end{figure}

The resulting lens redshift distribution for rays of type \IIorIII\ is illustrated in Fig.~\ref{fig:sigma_of_zL}, where the cross-section $\partial\tau^\mrm{I}_\IIorIII/\partial \zL$ is plotted as a function of lens redshift $\zL$ for various source redshifts $\zS$. (Not shown are the lens redshift distributions for rays with $|\mu|>10$ and with $r>10$, but these are very similar.) The inclusion of the stellar mass slightly increases the typical redshift of lenses (cf. fig. 4 in \citetalias{HilbertEtal2007_StrongLensing}), but most of the lenses still have $\zL<2.5$ even for high source redshifts. The lack of lenses at high redshift reflects the lower abundance of massive galaxies and halos, as well as the less favourable geometry for lensing at these redshifts.

\begin{figure}
\centerline{\includegraphics[width=1\linewidth]{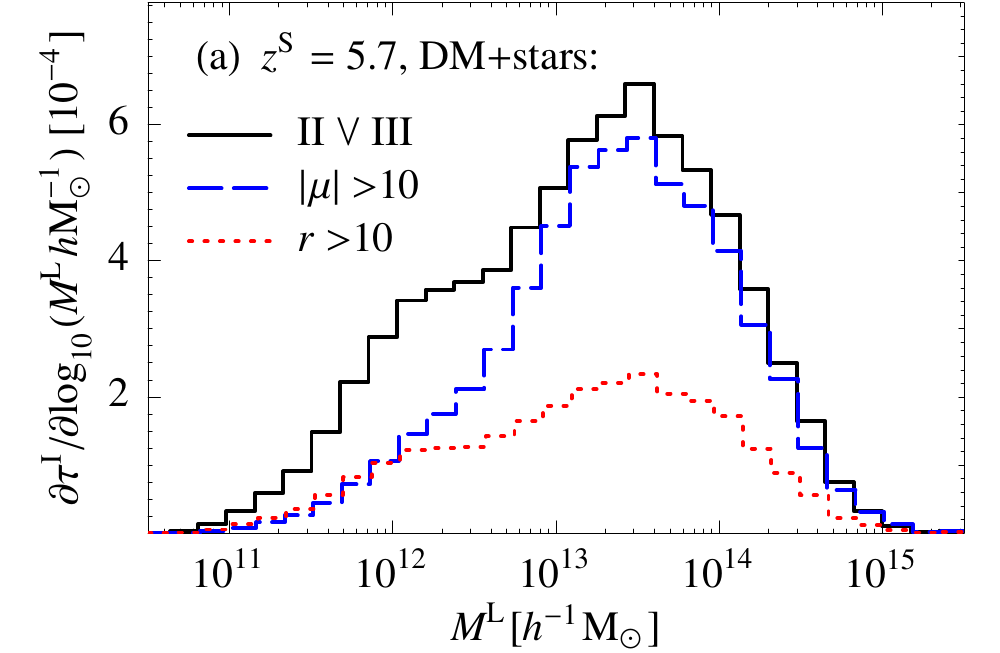}}
\centerline{\includegraphics[width=1\linewidth]{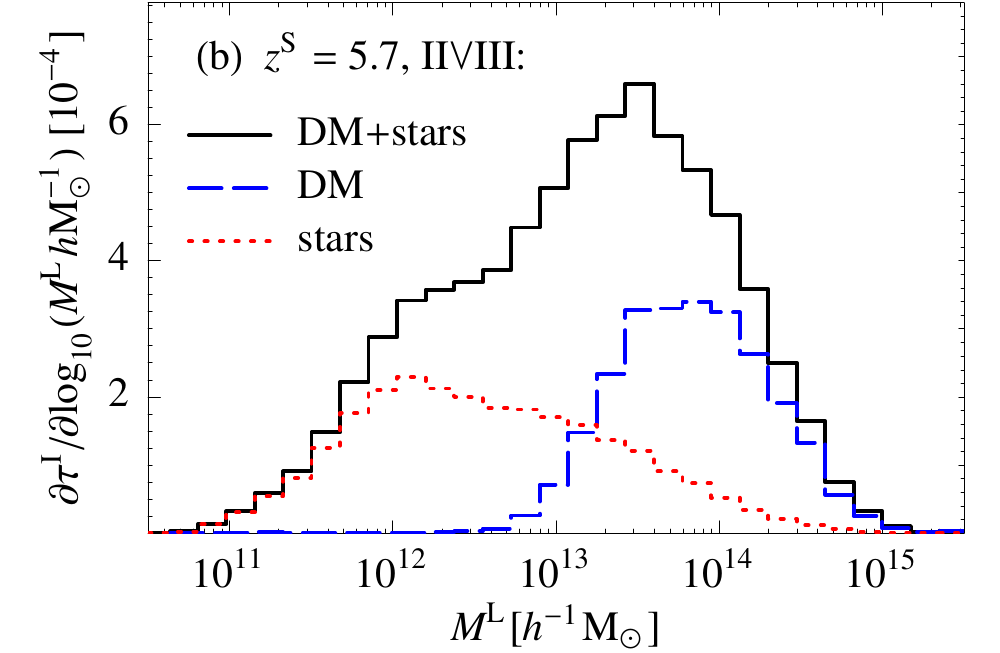}}
\caption{
\label{fig:sigma_of_logML}
The cross-section \mbox{$\partial\tau^\mrm{I}/\partial \log M^\mrm{L}$} as a function of the lens halo mass $M^\mrm{L}$ (see text) for sources at redshift $\zS=5.7$. Panel (a) compares rays of type \IIorIII\ (solid line), rays with \mbox{$|\mu|>10$} (dashed line), and rays with \mbox{$|r|>10$} (dotted line) for lensing by both dark and luminous matter. (b) compares rays of type \IIorIII\ for lensing by both dark and luminous matter (solid line), for lensing by dark matter alone (dashed line), and for lensing by luminous matter alone (dotted line).
}
\end{figure}

Following the method in \citetalias{HilbertEtal2007_StrongLensing}, we identify for each strongly lensed ray the dark-matter halo\footnote{The dark-matter halos considered here were identified in the simulation by applying a friend-of-friend group-finding algorithm to the dark-matter particle distribution.} associated with the lens by locating on the sufficient plane the halo with the highest ratio $M/b$ of its virial mass $M$ (defined as the mass within a sphere with mean enclosed density 200 times the cosmological mean) to the projected distance $b$ of its centre from the position where the ray intersects the plane. The resulting distributions of halo masses for rays of type \IIorIII, with $|\mu|>10$, and with $r>10$ are compared in Fig.~\ref{fig:sigma_of_logML}a, where the cross-sections $\partial\tau^\mrm{I}/\partial \log M^\mrm{L}$ are shown for $\zS=5.7$ as a function of lens halo mass $M^\mrm{L}$. For $M^\mrm{L}>10^{14}h^{-1}\,\Msolar$, the cross-sections increase by at most 40\% when the effects of the galaxies are included. For $M^\mrm{L}<10^{14}h^{-1}\,\Msolar$, the difference is much larger, however. In particular, the mass distribution of strongly lensing halos extends to significantly lower values when the galaxies are included (see Fig.~\ref{fig:sigma_of_logML}b). There is both a maximum at masses $M^\mrm{L}\approx3\times10^{13}h^{-1}\,\Msolar$ and a low-mass `tail'.

\begin{figure}
\centerline{\includegraphics[width=1\linewidth]{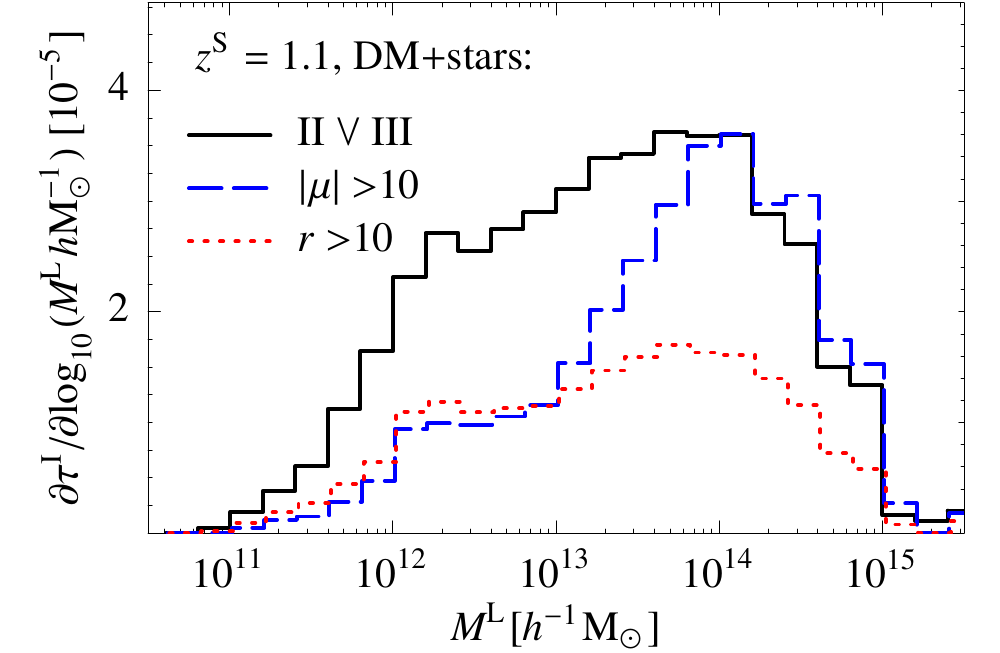}}
\caption{
\label{fig:sigma_of_logML_low_zS}
The cross-section \mbox{$\partial\tau^\mrm{I}/\partial \log M^\mrm{L}$} as a function of the lens halo mass $M^\mrm{L}$ for sources at redshift $\zS=1.1$. Compared are the rays of type \IIorIII\ (solid line), rays with \mbox{$|\mu|>10$} (dashed line), and rays with \mbox{$|r|>10$} (dotted line) for lensing by both dark and luminous matter.
}
\end{figure}

For lower source redshifts $\zS=1.1$, there is both a low-mass `tail' and a maximum at masses $M^\mrm{L}\approx 10^{14}h^{-1}\,\Msolar$ for rays with $|\mu|>10$ (see Fig.~\ref{fig:sigma_of_logML_low_zS}). For rays with $r>10$ and rays of \IIorIII, the distribution becomes bimodal with an additional maximum at $M^\mrm{L}\approx2\times10^{12}h^{-1}\,\Msolar$.

As can be seen in Fig.~\ref{fig:sigma_of_logML}a, $\partial\tau_{r>10}^\mrm{I}/\partial \log M^\mrm{L}$ and $\partial\tau_{|\mu|>10}^\mrm{I}/\partial \log M^\mrm{L}$ are very similar for $M^\mrm{L}\lesssim10^{12}h^{-1}\,\Msolar$. Strong lenses with these masses clearly show the lensing characteristics of isothermal spheres. Lenses with masses $M^\mrm{L}\gtrsim10^{13}h^{-1}\,\Msolar$, however, have much smaller cross-sections for $r>10$ than for $|\mu|>10$, which can be interpreted as `convergence-dominated' lensing, with shear being much less important than convergence.

For halos with mass $M^\mrm{L}<10^{13}h^{-1}\,\Msolar$, the dark matter alone is unable to produce strong lensing. However, the stellar mass of the galaxies changes this. As can be seen in Fig.~\ref{fig:sigma_of_logML}b, the cross-section is maximal for $M^\mrm{L}\approx10^{12}h^{-1}\,\Msolar$ if \emph{only} the \emph{stellar} mass is considered. Even though such galaxies typically have stellar masses $\Mstellar<10^{11}h^{-1}\,\Msolar$, and Einstein radii $\rEinstein\lesssim1\,\arcsect$, their high abundance apparently outweighs the small cross-sections of individual galaxies. Galaxies with such small Einstein radii produce strong lensing in our simulation because we use analytic expressions for the stellar contribution to the light deflection, which are not subject to the (larger) resolution limit of the meshes used for the the dark-matter contribution. One should, however, keep in mind that Einstein radii $\rEinstein\lesssim1\,\arcsect$ for galaxies with $\Mstellar<10^{11}h^{-1}\,\Msolar$ imply that our approach (which essentially assumes point-like sources) may overestimate the strong-lensing cross-sections for masses $M^\mrm{L}\lesssim10^{12}h^{-1}\,\Msolar$ and sources with comparable angular extent, e.g. distant galaxies.

The cross-sections for strong lensing by both dark and luminous matter shown in Fig.~\ref{fig:sigma_of_logML}a decrease rapidly for $M^\mrm{L}<10^{12}h^{-1}\,\Msolar$ and vanish for $M^\mrm{L}<5\times10^{10}h^{-1}\,\Msolar$. Apparently, halos with $M<5\times10^{10}h^{-1}\,\Msolar$ do not contribute to strong lensing even when the baryons in stars are taken into account.

\subsection{Images at larger radii}
\label{sec:large_image_splittings}
The inclusion of the stellar mass in galaxies greatly increases the optical depths for strong lensing. As discussed in the preceding section, a large part of the increase can be traced back to galaxies that have small Einstein radii $\rEinstein\lesssim1\,\arcsect$ and hence produce only small image splittings. This suggests that the effects of the luminous matter on the optical depths is smaller for larger image splittings.

To measure the impact parameter of a strongly lensed ray with respect to the lens centre -- and thus obtain an estimate for the expected image splitting -- we now consider the dark-matter subhalos\footnote{as identified by \mbox{\textsc{subfind}}~\citep{SpringelEtal2001_SUBFIND}} and -- in the case of lensing by both luminous and dark matter -- the galaxies\footnote{with their stars and their associated dark-matter subhalo (if there is one)} as individual lens candidates. For each strongly lensed ray, we identify the galaxy or dark-matter subhalo on the `sufficient' plane that has the largest ratio $M/b^2$ of its mass $M$ and the square of the projected distance between its centre and the ray. This ratio compares the impact parameter $b$ to the Einstein radius $r_\mrm{E}\propto\sqrt{M}$ of the lens candidate under the simplifying assumption of a point mass.\footnote{
Strongly lensed rays at larger impact parameters w.r.t. the halo centre are often lensed by substructure, which produce image splittings much smaller than the distance from the halo centre. The number of strongly lensed rays at large impact parameters w.r.t. the halo centre is therefore rather a measure of the amount of substructure present in dark-matter halos than a measure for large image splittings. If we identify the mass with the largest ratio $M/b$ as lens (as in the previous section), too often a cluster main halo is selected, although the lensing is due to a smaller non-central galaxy in that cluster (as is revealed by inspection of individual cases). For $M/b^2$, this problem does not occur.
}

\begin{figure}
\centerline{\includegraphics[width=1\linewidth]{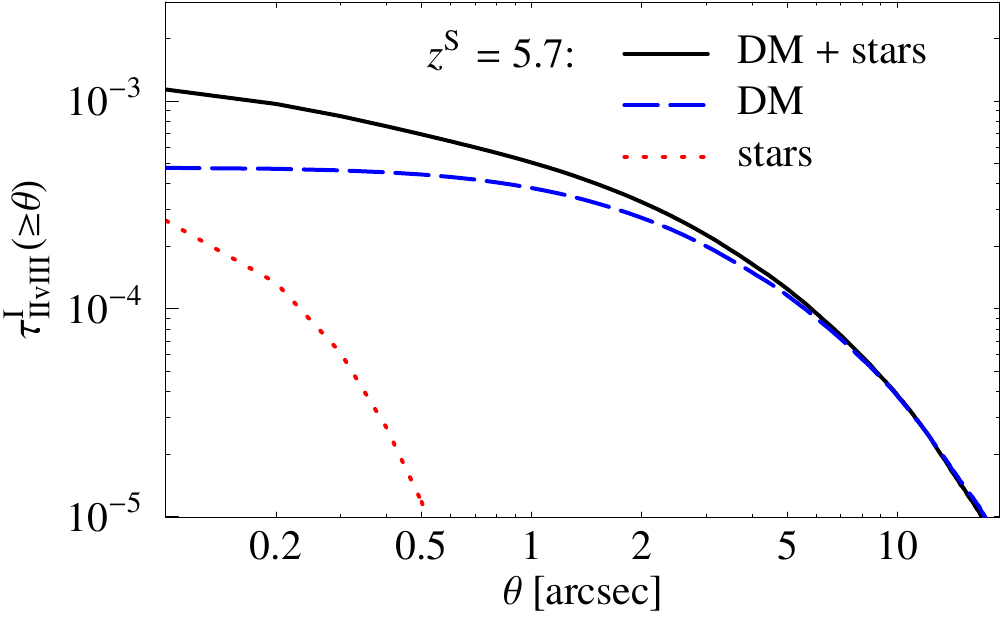}}
\caption{
\label{fig:tau_of_radius}
The optical depth $\tau^\mrm{I}_\IIorIII(\geq\theta)$ for rays of type \IIorIII\ with lens impact parameters $\geq\theta$. Compared are the depths for lensing by both dark and luminous matter (solid line), by dark matter alone (dashed line), and by luminous matter alone (dotted line) for sources at redshifts $\zS=5.7$.
}
\end{figure}

The resulting optical depths $\tau^\mrm{I}_\IIorIII(\geq\theta)$ for rays of type \IIorIII\ at angular separations $\geq\theta$ from the lens centre are plotted in Fig.~\ref{fig:tau_of_radius} for sources at redshift $\zS=5.7$. The difference between the optical depths for lensing by dark and luminous matter and lensing by dark matter alone are very large for images at radii $<1\,\arcsect$. However, the optical depths become more similar for larger radii so that the difference is less than $15$ per cent for radii $>5\,\arcsect$, and becomes negligible for radii $>10\,\arcsect$. The other optical depths for strong lensing exhibit a very similar behaviour as a function of minimal distance from the lens. These findings agree well with results of \citet{Oguri2006}, who studied the image-separation distribution using analytical mass profiles, or \citet{MeneghettiEtal2000}, \citet{MeneghettiBartelmannMoscardini2003} and \citet{DalalHolderHennawi2004}, who considered the influence of galaxies on the probabilities for giant arcs and arclets in clusters.
Furthermore, our results for the optical depths $\tau^\mrm{S}_{\mu>10}(\geq\theta)$ (not shown) agree well with the optical depths of \citet{WambsganssOstrikerBode2008}.

\begin{figure}
\centerline{\includegraphics[width=1\linewidth]{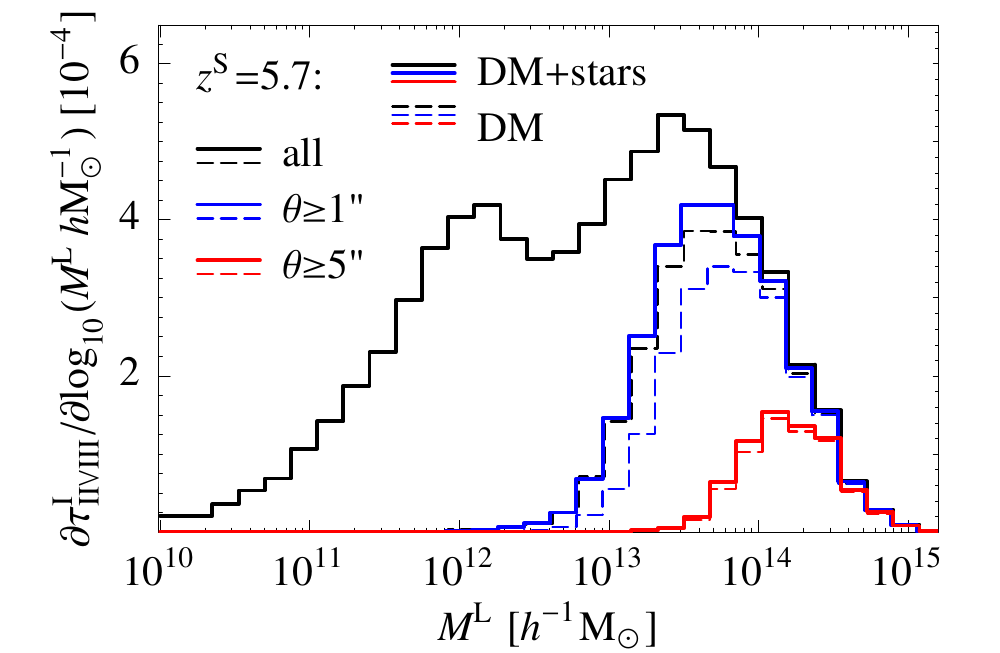}}
\caption{
\label{fig:sigma_of_logML_large_radii}
The cross-section \mbox{$\partial\tau^\mrm{I}_{\IIorIII,\geq\theta}/\partial \log M^\mrm{L}$} as a function of the lens halo mass $M^\mrm{L}$ (see text) for lensing by both dark and luminous matter (solid line) and lensing by dark matter (dashed line) for sources at redshift $\zS=5.7$. Compared are the cross-sections for rays of type \IIorIII\ regardless of their impact parameters (black lines), at impact parameters $\theta\geq1\,\arcsect$ (blue lines), and at $\theta\geq5\,\arcsect$ (red lines).
}
\end{figure}

The cross-section \mbox{$\partial\tau^\mrm{I}_{\IIorIII,\geq\theta}/\partial \log M^\mrm{L}$} for images of type \IIorIII\ at radii $\geq\theta$ is plotted in Fig.~\ref{fig:sigma_of_logML_large_radii} as a function of the lens halo mass $M^\mrm{L}$. If all images are considered, the cross-sections for lensing by both dark and luminous matter differs quite strongly from the cross-section for lensing by dark matter alone. Restricted to images at radii $\theta>1\,\arcsect$, however, the cross-sections become quite similar. In particular, the cross-section for strong lensing at radii $\theta>1\,\arcsect$ is very small for lenses with $M^\mrm{L}<10^{13}h^{-1}\,\Msolar $ even if the stellar mass is included. This confirms our assumption that the luminous and dark matter in smaller halos produces strongly lensed images only at radii $<1\,\arcsect$. For radii $\theta>5\,\arcsect$, there are virtually no lenses with masses below $M^\mrm{L}<2\times10^{13}\Msolar h^{1}$, and the cross-sections are almost identical.

\begin{figure}
\centerline{\includegraphics[width=1\linewidth]{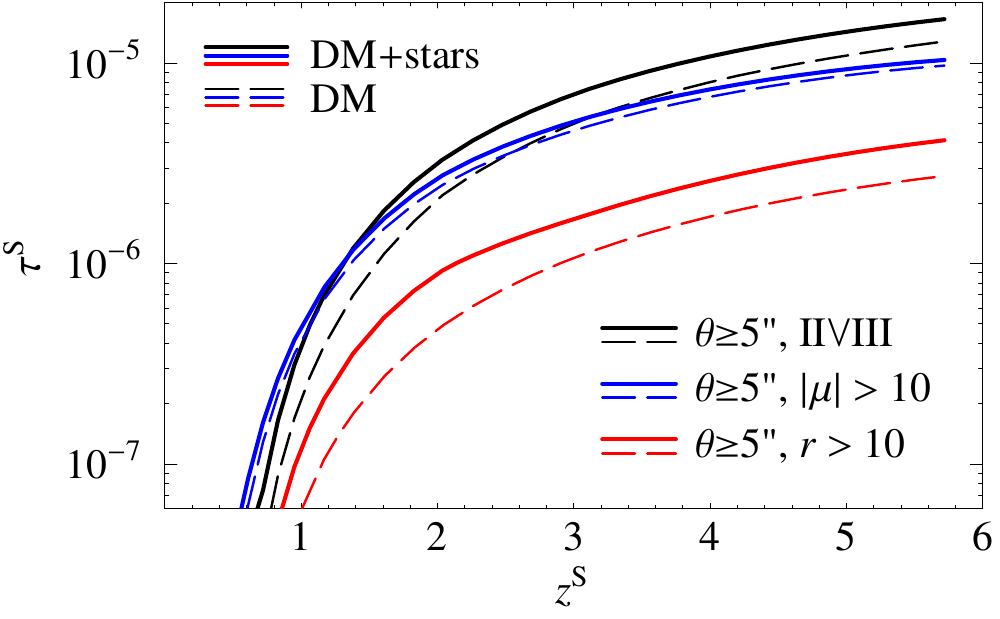}}
\caption{
\label{fig:tau_of_zL_large_radii}
The optical depth $\tau^\mrm{S}$ as a function of source redshift $\zS$ for images of small circular sources of type \IIorIII\ (black lines), with large magnification (blue) and with large length-to-width ratio (red) at lens impact parameters $\theta\geq5\,\arcsect$. Compared are the optical depths for lensing by both dark and luminous matter (solid lines) and by dark matter alone (dashed lines).
}
\end{figure}

The optical depths $\tau^\mrm{S}(\zS)$ for strong lensing by both dark and luminous matter as a function of source redshift $\zS$ are compared to the optical depths caused by dark matter alone in Fig.~\ref{fig:tau_of_zL_large_radii} for images with lens impact parameters $\geq5\,\arcsect$. These optical depths can be interpreted as being restricted to image splittings $\gtrsim10\,\arcsect$. The enhancement due to the stellar mass is larger for sources at lower redshift. For $\zS=1$, the stellar mass increases the optical depths by up to a factor two. For larger source redshifts the increase is $\lesssim1.5$ depending on source redshift and the property considered. Compared to the other optical depths, the optical depth for $|\mu|>10$ is affected least by the stellar matter. Note that the difference in the optical depths for lensing by both dark and luminous matter and by dark matter alone is larger for $\tau^\mrm{S}$ than for $\tau^\mrm{I}$ because of additional differences in the magnification distribution.

\subsection{Quasar lensing}
\label{sec:quasar_lensing}

The optical depths for images lensed by luminous and dark matter at radii $\geq5\,\arcsect$ (Fig.~\ref{fig:tau_of_zL_large_radii}) are an order of magnitude lower than the corresponding depths for images at radii $\geq0.5\,\arcsect$ (not shown). This ratio is consistent with results by \citet{InadaEtal2008}, who identified ten lensed quasars with image splittings $1$--$2\,\arcsect$ and one lensed quasar with a splitting of $15\,\arcsect$ in a statistical sample of $22\,683$ quasars with redshifts $0.6\leq z \leq 2.2$ from the Sloan Digital Sky Survey (SDSS).

The quasar sample of \citet{InadaEtal2008} has been used by \citet{OguriEtal2008} to constrain the matter content and dark-energy evolution of the Universe. \citet{HennawiDalalBode2007} estimated the number of expected number of quasars with very large image separations in similar samples. To obtain an estimate of the expected total number of lensed quasars in the sample of \citet{InadaEtal2008} from our data, we assume that every strongly lensed source with image splitting $\geq1\,\arcsect$ has one image of type II at radii $\geq0.5\,\arcsect$ from the lens centre. The quasar sample is limited to apparent $i$-band magnitudes $m_i<19.1$. For the calculation of the magnification bias, we use the quasar luminosity function obtained by \citet{RichardsEtal2005} from the 2dF-SDSS LRG and Quasar Survey and \citet{CroomEtal2004}:
\begin{equation}
\label{eq:Phi_g}
\Phi(L_g;z)=\frac{\Phi^*(z)}{L_g^*(z)}\left[\left(\frac{L_g}{L_g^*(z)}\right)^{-\alpha}+\left(\frac{L_g}{L_g^*(z)}\right)^{-\beta} \right]^{-1}.
\end{equation}
Here, $L_g$ denotes the rest-frame $g$-band luminosity, $z$ the quasar redshift, and $\Phi^*(z)$ a normalisation constant. The bright-end slope $\alpha=-3.31$, and the faint-end slope $\beta=-1.45$. The break luminosity $L_g^*(z)$ is parametrised as
\begin{equation}
L_g^*(z)=10^{k_1z+k_2z^2 -0.4 M_{g}^*(0)}L_{g0}
\end{equation}
with $M_{g}^*(0)=-21.61$, $k_1=1.39$, $k_2=-0.29$, and $L_{g0}$ as a luminosity standard.

We then convert the apparent $i$-band magnitude limit $m_i^\mrm{lim}=19.1$ to the corresponding $g$-band luminosity 
\begin{equation}
L_g^\mrm{lim}(z)=10^{-0.4\left[m_i^\mrm{lim}-\mrm{DM}(z)-K_{i,g}(z)\right]}L_0,
\end{equation}
where $\mrm{DM}(z)$ denotes the distance modulus to redshift $z$, and the $K$-correction $K_{i,g}(z)$ `corrects' between observer-frame $i$-band and rest-frame $g$-band. We use the $K$-correction discussed in \citet{RichardsEtal2006}, i.e. we assume $K_{i,g}=K_{i,i'}(z)-0.664$ with $K_{i,i'}(z)$ given by table~4 in \citet{RichardsEtal2006}.

The quasar lensing cross-section $\sigma_\mrm{QL}(z)$ as a function of redshift $z$ is then calculated from our ray sample by \citep{SchneiderEhlersFalco_book}:
\begin{equation}
\label{eq:quasar_lensing_cross_section}
\begin{split}
\sigma_\mrm{QL}(z)&=\frac{1}{N_\mrm{rays}}\sum_{i=1}^{N_\mrm{rays}} |\mu(i;z)|^{-1} 1_\mrm{QL}{(i;z)}
\times\\&\hspace{3em}\times
\frac{\int_{|\mu(i;z)|^{-1}L_g^\mrm{lim}(z)}^{\infty} \Phi(L_g;z)\diff{L_g}}
{\int_{L_g^\mrm{lim}(z)}^{\infty}\Phi(L_g;z)\diff{L_g} },
\\&=
\frac{1}{N_\mrm{rays}}\sum_{i=1}^{N_\mrm{rays}} |\mu(i;z)|^{-1} 1_\mrm{QL}{(i;z)}
\times\\&\hspace{6em}\times
\frac{F(|\mu(i;z)|^{-1}x^\mrm{lim}(z))} {F(x^\mrm{lim}(z))}.
\end{split}
\end{equation}
Here, $N_\mrm{rays}$ denotes the total number of rays, $\mu(i;z)$ denotes the magnification of ray $i$ for sources at redshift $z$, and $1_\mrm{QL}{(i;z)}=1$ if ray $i$ is of type II with lens impact parameter $\geq0.5\,\arcsect$, and zero otherwise,
\begin{equation}
F(x^\mrm{lim})= \int_{x^\mrm{lim}}^\infty\left[x^{-\alpha}+x^{-\beta} \right]^{-1}\diff{x},
\end{equation}
and $x^\mrm{lim}(z)=L_g^\mrm{lim}(z)/L_g^*(z)$. The factor $|\mu(i;z)|^{-1}$ in Eq.~\eqref{eq:quasar_lensing_cross_section} accounts for the fact that we assume a uniform distribution of sources in the source plane, while our ray sampling method uniformly samples the image plane. The ratio of the cumulative luminosity distributions quantifies the increase of source counts due to a lower luminosity threshold for detection in regions of higher magnification.

Under the assumption that the total number of quasars is not affected by magnification bias, the number of lensed quasars $N_\mrm{QL}$ in the sample of \citet{InadaEtal2008} is obtained by integrating over the observed redshift distribution $n_\mrm{Q}(z)$ of the quasar sample\footnote{http://www-utap.phys.s.u-tokyo.ac.jp/\textasciitilde{}sdss/sqls/} weighted by the cross-section:
\begin{equation}
N_\mrm{QL}=\int_0^\infty n_\mrm{Q}(z) \sigma_\mrm{QL}(z) \diff{z}.
\end{equation}

If we consider lensing by dark matter alone, we predict less than one lensed quasar with image splitting $\geq1\,\arcsect$. For lensing by both luminous and dark matter, three lensed quasars with image splitting $\geq1\,\arcsect$ are predicted. This is still small compared to the 11 lensed quasars in the sample. One reason for this discrepancy could be that our simple ray-selection criterion $1_\mrm{QL}$ does not provide a very good approximation for the fraction of strongly lensed sources with image splittings $>1\,\arcsect$.\footnote{The criterion $1_\mrm{QL}$ effectively excludes (the abundant) lenses with $\rEinstein<0.5\,\arcsect$, which do not produce image splittings $\gtrsim 0.5\,\arcsect$. Undesirably, the criterion also excludes those multi-image systems with image separations $>0.5\,\arcsect$ where the type-II image is near the centre of the lens.} Furthermore, we certainly underestimated the magnification bias.\footnote{The magnitude limit for the quasar sample is $i_\mrm{cor}\leq19.1$, where $i_\mrm{cor}$ is the Galactic-extinction corrected magnitude of the brightest image with an image-separation dependent admixture of the fainter image(s) and the lens galaxy \citep{OguriEtal2006}. Moreover, the brightest image is often the primary image, which is of type I, but \emph{not} of type II.} Other possible reasons for the discrepancy may be effects of non-spherical baryon distribution or halo contraction due to baryon condensation. These are not considered here. Imitating the contraction effect by simply doubling the stellar mass in the galaxies, we predict about 8 lensed quasars, much closer to the observed numbers of lensed quasars. These results indicate that predictions are very sensitive to detailed assumptions about galaxy and dark matter structure on small scales, and are thus unlikely to provide robust constraints on cosmological parameters.

\section{Summary}
\label{sec:summary}
In this work, we have studied how the stellar components of galaxies affect predictions for gravitational lensing in the concordance $\Lambda$CDM cosmology. Our results were obtained by shooting random rays though a series of lens planes created from the Millennium Simulation. The dark-matter component on the lens planes was constructed directly from the dark-matter particle distribution of the simulation, while the distribution of the luminous matter was obtained from semi-analytic galaxy formation models run on stored merger trees from the simulation.

In Sec.~\ref{sec:magnification_distribution} we discussed the influence of stellar mass on the statistical distribution of point-source magnifications. Although this distribution is almost unchanged for magnifications $\mu\approx1$, the galaxies induce a noticeable increase of the probability for very high and very low magnifications.

In Sec.~\ref{sec:optical_depths}, we presented optical depths for images of small sources that are highly magnified, strongly distorted or belong to multiply imaged sources. We compared the results obtained by including both dark and luminous matter to those obtained for dark matter alone. We find that the inclusion of the luminous matter greatly enhances the strong-lensing optical depths compared to a `dark-matter only' universe. Our results in Sec.~\ref{sec:lens_properties} show that the increase is partly due to a significant enhancement of the strong-lensing cross-section of group and cluster halos with virial masses $M>10^{13}h^{-1}\,\Msolar$. In addition, the stellar matter leads to significant strong lensing in smaller halos, which would not cause noticeable strong lensing otherwise. Although these halos have typical Einstein radii $\rEinstein\lesssim1\,\arcsect$, their large abundance outweighs their small individual cross-sections, leading to a bimodal distribution of integrated cross-sections with halo mass both for strongly distorted images and for multiply imaged sources at redshifts $\zS\lesssim1$.

In Sec.~\ref{sec:large_image_splittings}, we estimated optical depths for strongly lensed images formed at larger distance from centre of their lens. Although the optical depths for lensing by both the dark and luminous matter are much larger than for lensing by dark matter alone for images at radii $<1\,\arcsect$, for images at larger radii the optical depths are much more similar. For images at radii $\geq5\,\arcsect$, the optical depths differ by at most a factor of two. For radii $\gtrsim10\,\arcsect$, there is almost no enhancement due to the galaxies, in agreement with earlier studies \citep[e.g.][]{MeneghettiBartelmannMoscardini2003,DalalHolderHennawi2004}.

Our results are consistent with the observed radial distribution of multiply imaged quasars with splittings $\geq1\,\arcsect$ in the recent SDSS quasar sample of \citet{InadaEtal2008} {\it only} if the effect of the galaxies is taken into account. The total number of lensed quasars predicted by our standard model is still quite low compared to the number observed, indicating a need to include the effects of baryonic dissipation on the dark matter distribution in order to explain the data fully.

An obvious extension to the work presented here will be ray-tracing over finite fields to study the effects of lensing on sources with finite extent. With a realistic distribution of source properties, the results could be compared directly with observations of massive galaxy clusters, where many distorted images and multi-image systems have been studied in some detail \citep{BroadhurstEtal2005,HalkolaSeitzPannella2006}.

Here, we used only the stellar mass and the size information from the semi-analytic galaxy formation models. In future work, the morphology and luminosity information could be added to simulate galaxy-galaxy lensing surveys with given selection functions. This will also allow us to take into account the effects of the light of lens galaxies on the detectability of strongly lensed images near the lens centre \citep{MeneghettiEtal2008}. With some improvements in the modelling of the galaxies, in particular by assuming realistic elliptical mass and light profiles for the stellar components, such simulated surveys will provide detailed predictions for galaxy-galaxy lensing.

The dark-matter distribution in galaxies, groups and clusters does not merely provide an arena for physical processes in the baryonic gas, it is also subject to modification by these processes \citep{BlumenthalEtal1986,GnedinEtal2004}. Baryon condensation increases the dark-matter density in the inner parts of halos, and so affects their dark-matter lensing properties \citep{PuchweinEtal2005,JingEtal2006,RozoEtal2006_astroph}. The semi-analytic models used here do not yet describe these processes, but they should be included in future work aiming at higher precision.

\section*{Acknowledgments}
We thank Volker Springel, Gabriella De Lucia, J{\'er}{\'e}my Blaizot, Manfred Kitzbichler, and Ole M{\"o}ller for helpful discussions. This work was supported by the DFG within the Priority Programme 1177 under the projects SCHN 342/6 and WH 6/3.

\appendix


\end{document}